\def\ang{\AA}
\def\gapprox{\lower.4ex\hbox{$\;\buildrel >\over{\scriptstyle\sim}\;$}}
\def\lapprox{\lower.4ex\hbox{$\;\buildrel <\over{\scriptstyle\sim}\;$}}

\def\refer#1   {\par\noindent\hangindent1cm {#1}}

 \documentstyle[11pt,aaspp4]{article}   
 \input epsf
\slugcomment{{\sl The Astrophysical Journal Letter }   \hfill{Revised: 2002 March 24}}

\begin{document}

\title{         A NEW METHOD TO CONSTRAIN THE IRON ABUNDANCE
		FROM COOLING DELAYS IN CORONAL LOOPS 			}
  
\author{Markus J. Aschwanden\footnote{Lockheed Martin Advanced Technology Center,
                Solar \& Astrophysics Laboratory,
                Dept. L9-41, Bldg.252,
                3251 Hanover St.,
                Palo Alto, CA 94304, USA;
                e-mail: aschwanden@lmsal.com},
	Carolus J. Schrijver$^1$,
	Amy R. Winebarger\footnote{Computational Physics, Inc., 8001 Braddock Road, 
		Suite 210, Springfield, VA 22151}$^{,}$\footnote{Naval Research Laboratory
		Code 7673, Washington DC 20375, USA}, 
 	and Harry P. Warren$^3$ }

\begin{abstract}
Recent observations with TRACE reveal that the time delay between
the appearance of a cooling loop in different EUV temperature filters is
proportional to the loop length, $\Delta t_{12} \propto L$. We model 
this cooling delay in terms of radiative loss and confirm this linear
relationship theoretically. We derive an expression that can be used to
constrain the coronal iron enhancement ${\alpha}_{Fe}=A_{Fe}^{cor}/A_{Fe}^{Ph}$ 
relative to the photospheric value as function of
the cooling delay $\Delta t_{12}$, flux $F_2$, loop width $w$, and
filling factor $q_w \le 1$. With this relation we find upper limits on the
iron abundance enhancement of ${\alpha}_{Fe} \le 4.8\pm 1.7$ for 10
small-scale nanoflare loops, and ${\alpha}_{Fe} \le 1.4\pm 0.4$ for 5
large-scale loops, in the temperature range of $T\approx 1.0-1.4$ MK. 
This result supports the previous finding that low-FIP elements, including Fe, 
are enhanced in the corona. The same relation constitutes also a lower limit 
for the filling factor, which is $q_w \ge 0.2\pm 0.1$ and 
$q_w \ge 0.8\pm 0.2$ for the two groups of coronal loops.
\end{abstract}

\keywords{Sun: Corona --- Sun: UV radiation }

\section{              Introduction 			}

Coronal EUV emission is mainly produced by radiative decay of collisionally
excited, highly-ionized iron ions, i.e. by Fe$^{8+}$ and Fe$^{9+}$ in the 171 \ang\ 
TRACE passband ($T\approx 1.0$ MK) and by Fe$^{11+}$ at 195 \ang\ ($T \approx 1.5$ MK).
Coronal loops undergo various phases of heating and cooling. When steady heating 
operates, the loops evolve into a steady-state, where heating input is
balanced by thermal conduction losses to the chromosphere and radiative 
losses into space, as described by the energy equation derived by
Rosner, Tucker, \& Vaiana (1978), generalized for gravity and non-uniform 
heating by Serio et al. (1981). When heating stops, coronal loops 
cool off by thermal conduction and radiative losses. One consequence of
this cooling process is that the EUV emission peaks first in a 
high-temperature filter 1, e.g. in TRACE 195 \ang , and later in a
lower-temperature filter 2, e.g. in TRACE 171 \ang\ , with a time delay
that we call {\sl cooling delay} $\Delta t_{12}$. This cooling delay
was found to scale proportionally to the loop length $L$ in a recent
study (Winebarger et al. 2003). Under the assumption that the cooling
time is dominated by radiative cooling, we can relate this observable
time delay to the radiative loss function $\Lambda (T)$, which allows
us to constrain the absolute abundances of iron ions, which dominates the
radiative loss function in this coronal temperature range of $T\approx
1-2$ MK. The new method of determining iron abundances provides an 
important diagnostic for coronal heating mechanisms that involve preferential 
ion heating. This study demonstrates also that much more physical information 
can be inferred from the temporal evolution of the EUV intensity than from 
the intensity measured in a single image.

\section{ 		Theoretical Model 				}

Plasma cooling through a narrow temperature range can be approximated
by an exponential function,
\begin{equation}
	T_e(t) = T_1 \exp{\left( - {t \over {\tau}_{cool}} \right)} \ ,
\end{equation}
where $T_1$ is the initial temperature at time $t=0$ and ${\tau}_{cool}$
is the cooling time. Cooling over large temperature ranges would require
the full consideration of the hydrodynamic equations, but the exponential
approximation is fully justified for the narrow temperature range of
$T_e \approx 1.0-1.4$ MK we are considering here, during the cooling of
coronal flare loops through the two TRACE 171 and 195 \ang\ filters
(see also measurements in Warren et al. 2003). 
So, when a cooling plasma reaches the temperature $T_1$ of the peak
response of the hotter filter ($T_1=1.4$ MK for TRACE 195 \ang ),
the time delay $\Delta t_{12}$ to cool down to the cooler 
filter $T_2$ (e.g. $T_2=0.96$
MK for 171 TRACE \ang ), can be expressed with Eq.1 as
\begin{equation}
	\Delta t_{12} = t_2 - t_1 = {\tau}_{cool} \ln{\left( {T_1 \over T_2} \right)} \ .
\end{equation}
The cooling time scale could be dominated by thermal conduction losses
in the initial phase, but is always dominated 
by radiative losses in the later phase (e.g., Antiochos \& Sturrock 1982).
Here we make the a priori assumption that the cooling
of EUV loops in the $T_e \approx 1$ MK temperature range is dominated
by radiative losses. This working hypothesis is particularly justified
near the almost isothermal loop tops and is also corroborated by the 
observational result found in Aschwanden et al. (2000), where a median value of 
${\tau}_{cool}/{\tau}_{rad}=1.02$ was obtained from the statistics of
12 nanoflare loops observed with TRACE 171 and 195 \ang . 
Moreover we show later that the radiative cooling time is always significantly
shorter than the conductive cooling time for the cases analyzed here. Thus we set
the cooling time ${\tau}_{cool}$ equal to the radiative cooling time
${\tau}_{rad}$,
\begin{equation}
	{\tau}_{cool} \approx 
        {\tau}_{rad} = {E_{th} \over dE_R/dt} = {3 n_e k_B T_e \over n_e n_H {\alpha}_{FIP} \Lambda (T_e)}
\end{equation}
where $n_e$ is the electron density, $n_H$ the hydrogen density,
$T_e$ the electron temperature,
$k_B=1.38 \times 10^{-16}$ erg K$^{-1}$ the Boltzmann constant, 
${\alpha}_{FIP}$ the abundance enhancement factor for low 
{\sl first-ionization potential} elements (at $< 10$ keV), and 
$\Lambda (T_e)$ the radiative loss function, which can be approximated
with a constant in the limited temperature range of $T_e\approx 0.5-2.0$
MK, according to the piece-wise powerlaw approximation of Rosner et al. (1978), 
$	\Lambda (T) \approx {\Lambda}_0 = 10^{-21.94}\ {\rm erg}\ {\rm s}^{-1}\ {\rm cm}^3$,
for T=0.5-2.0 MK.
The computation of the radiative loss function at a given temperature depends on 
the elemental atomic abundances,
and thus we define a reference value ${\Lambda}_{0,Ph}$ for photospheric abundances.
Coronal abundances generally show a density enhancement for low first-ionization
potential (FIP) elements, which we express with an enhancement factor ${\alpha}_{FIP}$.
Since iron (Fe) contribution strongly at these temperatures, the radiative loss function 
in the corona scales as ${\Lambda}_{0,cor} = {\Lambda}_{0,ch} \times {\alpha}_{Fe}$. 
The cooling delay $\Delta t_{12}$ as function of the coronal iron abundance ${\alpha}_{Fe}$ 
is thus (with Eqs.~1-3), assuming full ionization in the corona ($n_H = n_e$),
\begin{equation}
        \Delta t_{12} = {3 k_B T_e \over n_e {\alpha}_{Fe} {\Lambda}_{0,Ph}} 
	\ln{\left( {T_1 \over T_2} \right) } \ .
\end{equation}
When a loop cools through a passband, the maximum of the flux $F(t)$ is detected at the
time when the loop temperature matches the peak of the response function, 
so the peak flux $F_2$ of the light
curve in the lower filter corresponds to the emission measure $EM$ at the
filter temperature $T_2$,
\begin{equation}
	F_2 = EM \times R_2 = (n_e^2 w q_w) \times R_2
\end{equation}
with the flux $F_2$ in units of DN/(pixel s), $w$ is the loop width or diameter, 
$q_w$ is the linear filling factor in case of unresolved substructures, and
$R_2$ is the response function, which is
$R_2=0.37 \times 10^{-26}$ cm$^{5}$ DN/(pixel s) for 171 \ang\ and photospheric
abundances (see Appendix A in Aschwanden et al. 2000). The value $R_2$ of the
response function refers to the time at the beginning of the mission, while
the degradation decreased this value by a factor of 0.78 in August 1999, the
latest date of analyzed observations used here. The corresponding correction
by a factor of $\ge \sqrt{0.78}=0.88$ is neglected in the numerical values given in Table 1.
Inserting the density from Eq.(5) into Eq.(4) we find the following expression
for the iron abundance ${\alpha}_{Fe}$, 
\begin{equation}
	{\alpha}_{Fe} = {3 k_B T_2 \over {\Lambda}_{0,Ph} \Delta t_{12}} 
			\sqrt{{R_2 w q_w \over F_2}}
			\ln{\left({T_1\over T_2}\right)} 
			= 4.17 \left({w \ q_w \over 1\ {\rm Mm}}\right)^{1/2}
			       \left({F_2 \over 10\ {\rm DN/s}}\right)^{-1/2}
			       \left({\Delta t_{12} \over 1\ {\rm min}}\right)^{-1} \ .
\end{equation}
For a filling factor of unity ($q_w=1$), the iron enhancement factor can be determined
with an accurcay of about $\lapprox 20$\%, because the observables $w$, $F_2$, and
$\Delta t_{12}$ can each be measured better than $\lapprox 10$\%. In case of
unresolved fine structure, i.e. filling factors of $q_w < 1$, we obtain with Eq.(6)
an upper limit for the iron enhancement. 

\medskip 
Recent EUV observations with TRACE have shown that the cooling delay $\Delta t_{12}$
is roughly proportional to the loop length $L$. In order to understand such a correlation
we use the energy balance equation, which is valid in a steady-state, e.g., 
before the cooling process, at the turning point from dominant heating
to dominant cooling, or at the turning point from dominant conductive cooling
to radiative cooling. The resulting scaling law is according to 
Rosner et al. (1978),
\begin{equation}
	T_{max} \approx 1400 ( p_0 L )^{1/3} \times q_{Serio} 
\end{equation}
with $p_0$ the pressure and $L$ the loop half length.
This scaling law has been generalized for gravity and non-uniform heating 
by Serio et al. (1981), modified by the correction factor
\begin{equation}
	q_{Serio} = \exp{\left( -0.08 {L \over s_H} - 0.04 {L \over s_p} \right)} \ ,
\end{equation}
where $s_H$ is the heating scale length and $s_p=47,000 \times T_{MK}$ km the pressure scale height.
With the ideal gas law ($p = 2 n_e k_B T_{max} = p_0 q_p$), corrected for the pressure at the loop
top [$q_p = \exp{( - h / s_p)} = \exp{( - {2 L / \pi s_p} )}$], we can eliminate the pressure $p$
in the RTV scaling law and find the following expression for the density $n_e$,
\begin{equation}
	n_e = {T_{max}^2 \over 2 k_B L (1400\ q_{Serio})^3 q_p } \ .
\end{equation}
Inserting this density into the relation for the cooling delay (Eq.~4) we find
indeed a proportional relation $\Delta t_{12} \propto L$, 
\begin{equation}
	\Delta t_{12} = L \times \left[ 
	{ 6 \ (1400\ q_{Serio})^3 q_p \ k_B^2 \over T_{max} {\Lambda}_{0,Ph} {\alpha}_{Fe} } 
	\ln{\left( {T_1 \over T_2} \right) } \right] \ .
\end{equation}
which should show up for cooling loops with similar maximum temperatures $T_{max}$.
The relation ${\tau}_{cool} \propto L$ was also derived in
Cargill et al. (1995; Eq.~14E therein) and Serio et al. (1991; Eq.~14 therein).

\section{	Data Analysis		}

We are using three data sets for which the cooling delay between the TRACE
195 \ang\ and 171 \ang\ filters has been measured:
11 nanoflare loops analyzed in Aschwanden et al. (2000), 
4 medium-sized EUV loops analyzed in Schrijver (2002), and 
5 large EUV loops analyzed in Winebarger at al. (2003). 

In the first study (Aschwanden et al. 2000), nanoflare loops were measured on 
1999 Febr 17, 02:15-03:00 UT, with a cadence of $\approx 2$ min in both
171 \ang\ and 195 \ang\ with TRACE. The time delay between the appearance
in the two filters was measured by cross-correlation of the two time profiles,
shown in Fig.6 and listed in Table 1 of Aschwanden et al. (2000). The time
delay was found to be positive in 11 out of 12 cases. We use only the 11
cases with positive time delay, sorted according to the loop length in Table 1.
The loop length was determined from a geometric model of a projected 
semi-circular cylindrical loop. The electron density $n_e$ is measured from 
the flux $F_2$ and loop width $w$ according to Eq.(5). 

In the second study (Schrijver 2001), time delays between the maximum intensity
in the 171 and 195 \ang\ TRACE passbands were measured in 4 active region loops
observed above the limb on 2000 May 26. The loop half length was estimated with 
$L=(\pi / 2) h + h_{limb}$ based on the height $h$ of the loop top above the limb
and a height correction $h_{limb}\approx 5$ Mm for the offset between 
optical limb and the portion of the plage inside the limb. The fluxes $F_{171}$ and
loop widths $w$ could not be reliably measured for this subset, because of confusion
problems in the crowded limb regions.

In the third study (Winebarger et al. 2002), 5 loops were measured on
1999 Aug 18, 1998 Jul 04, 1998 Jul 25,
1998 Aug 17, and 1998 Jul 25, in both the 171 \ang\
and 195 \ang\ filters with TRACE. The time delay between the two filters was measured
from the peak times of asymmetric gaussian curves fitted to the light 
curves in the two filters. The loop lengths were measured from the best
fit of elliptical and dipolar geometric models to the projected loop shapes. 

Inserting the measured values $F_2$, $w$, and $\Delta t_{12}$ into Eq.(6) 
yields the iron abundance enhancement factors ${\alpha}_{Fe}$ listed in Table 1, 
which have a mean and standard deviation of
${\alpha}_{Fe} \le 4.9 \pm 1.7$ for the dataset of Aschwanden et al. (2000), and
${\alpha}_{Fe} \le 1.4 \pm 0.4$ for the dataset of Winebarger et al. (2003), respectively.
Note that the mean iron enhancement is significantly higher for the small-scale
nanoflares analyzed in Aschwanden et al. (2000) than for the large-scale loops
of Winebarger et al. (2003). The variable degree of iron enhancement could be related 
to different physical conditions in freshly-filled small-scale loops compared
with longer-lived large-scale loops. The effect of gravitational settling has
been observed in observations of coronal streamers with SUMER (Feldman et al.
1999) and with UVCS (Raymond et al. 1997).

The values of ${\alpha}_{Fe}$ have to be considered as upper limits if the
filling factor $q_w \le 1$ is lower than unity. We can turn the argument around
and assume that the iron abundance enhancement has to be larger than or equal unity,
which would then constitute lower limits for the filling factors:
$q_w \ge 0.23 \pm 0.08$ (Aschwanden et al. (2000) and
$q_w \ge 0.78 \pm 0.22$ (Winebarger et al. (2003).

Figure 1 shows the correlation plot between the cooling delay ${\Delta}t_{12}$
and the loop half length $L$. A linear regression fit between the logarithic
values yields the power-law relation ${\Delta}t_{12} \propto L^{1.08\pm0.16}$
which is fully consistent with the theoretical prediction of a linear
relationship ${\Delta}t_{12} \propto L^1$ (Eq.~10). 

In order to verify our initial assumption of dominant radiative loss, we estimate
also the conductive cooling time,
\begin{equation}
	{\tau}_{cond} = {3 n_e k_B T_{max} \over \nabla F_C} =
			1.1 \times 10^{-9} n_e T_{max}^{-5/2} L^2 \ ,
\end{equation} 
where we assign a mean value of $T_{max} \approx 1.2$ MK for the looptop temperature,
when the loop cools through the two TRACE passbands. The last two
columns in Table 1 show that the conductive cooling time is always much larger than
the radiative cooling time, ${\tau}_{cond} \gg {\tau}_{rad}$, 
which corroborates our a priori assumption of dominant radiative cooling. 

\section{	Discussion	}

We have developed a simple method to constrain the iron abundance or filling factor in 
a coronal loop, based on the cooling delay measurement between two EUV filters. 
This model predicts a linear relationship between
the cooling delay and the loop length, i.e. ${\Delta}t_{12} \propto L$, which
is consistent with the observed relation, i.e. ${\Delta}t_{12} \propto L^{1.08\pm0.16}$.

The only underlying assumption is that the cooling is dominated by radiative loss
in the temperature range of EUV loops ($T\approx 1.0-1.4$ MK here), rather than by 
conductive loss. Theoretical models of cooling in flare loops predict that conductive 
cooling is only dominant in the initial phase of very hot plasma seen in soft X-rays, 
say at $T \gapprox 10$ MK, while the later cooling phase seen in EUV is dominanted 
by radiative cooling (Antiochos 1980; Antiochos \& Sturrock 1982; Cargill et al. 1995). 
Our measurements of the 
density allows us to estimate upper limits for the radiative cooling time. If there 
would be a filling factor $q_w < 1$, the density would be higher and the radiative 
cooling time shorter. But even for a filling factor of unity,
we find that the radiative cooling time in the EUV temperature range
is significantly shorter than the conductive cooling time, which corroborates our 
a priori assumption. This is also consistent with other observations of EUV loops, 
where the ratio of the cooling time to the radiative cooling time was found to be 
${\tau}_{cool}/{\tau}_{rad}=1.02$ (Aschwanden et al. 2000).

The main result of this study is the estimation of the iron abundance. We
show a compilation of radiative loss functions in Fig.~2, which has been calculated
for photospheric abundances (Meyer 1985), with an absolute iron abundance of 
$log(A_{Fe})=7.59$ relative to hydrogen $log(A_H)=12.0$ 
(i.e. $A_{Fe}/A_H=3.9\times 10^{-5}$), as well as for
coronal abundances (Feldman 1992), which have an iron enhancement by a factor of ${\alpha}_{Fe}=3$.
This enhancement factor of ${\alpha}_{Fe}=3$ in density produces a change of the radiative
loss rate that can be seen between the two curves calculated by Martens et al. (2000) in Fig.~2,
at a temperature of $T\approx 1.0$ MK. Recent measurements of the absolute abundance of
iron based on comparisons or EUV and radio data yielded a value of $A_{Fe}/A_H=1.56 \times
10^{-4}$, or a coronal iron enhancement by a factor of ${\alpha}_{Fe}=4.0$ (White et al. 2000).    
Our measurements from 16 different loops in many different active regions yield
a median value of ${\alpha}_{Fe} = 4.0$, or a mean and standard deviation of
${\alpha}_{Fe} = 3.7\pm2.2$. Because of this reasonable agreement of iron enhancements 
with radio methods (White et al. 2000) and spectroscopic measurements (Feldman 1992), 
our result corroborates the notion that 
low-FIP elements such as Fe are enhanced in the corona relative to photospheric values 
(Feldman 1992). By the same token we can argue the filling factor is close to unity
in coronal EUV loops, otherwise we would have disagreement with spectroscopic and radio 
iron abundance measurements.

\acknowledgements
{\sl Acknowledgements:}
Part of this work was supported by NASA contracts NAS5-38099 (TRACE) 
and NAS8-00119 (SXT).

\section*{               References                                              }

\refer{Antiochos,S.K. 1980, ApJ 241, 385}
\refer{Antiochos,S.K. \& Sturrock,P.A. 1982, ApJ 254, 343.}
\refer{Aschwanden,M.J., Tarbell,T.T., Nightingale,W., Schrijver,C.J., Title,A.,
	Kankelborg,C.C., Martens,P., and Warren,H.P. 2000, ApJ 535, 1047}
\refer{Aschwanden,M.J. and Schrijver,C.J. 2002, ApJS, 142, 269}
\refer{Cargill,P.J., Mariska,J.T., and Antiochos,S.K. 1995, ApJ 439, 1034}
\refer{Cook,J.W., Cheng,C.C., Jacobs,V.L., and Antiochos,S.K. 1989, ApJ 338, 1176}
\refer{Feldman,U. 1992, Physica Scripta 46, 202}
\refer{Feldman,U., Doschek,G.A., Sch\"uhle,U., and Wilhelm,K. 1999, ApJ 518, 500.}
\refer{Golub,L. and Pasachoff,J.M. 1997, {\sl The Solar Corona}, Cambridge: Cambridge University Press}
\refer{Martens,P.C.H., Kankelborg,C.C., and Berger,T.E. 2000, ApJ 537, 471}
\refer{Meyer,J.-P. 1985, ApJS 57, 173}
\refer{Raymond,J.C. and 26 co-authors 1997, Solar Phys. 170, 105}, 
\refer{Rosner,R., Tucker,W.H., and Vaiana,G.S. 1978, ApJ 220, 643}
\refer{Schrijver,C.J. 2001, Solar Phys. 198, 325} 
\refer{Schrijver,C.J. and Zwaan,C. 2000, {\sl Solar and stellar magnetic activity},
 		2000, Cambridge: Cambridge University Press}
\refer{Serio,S., Peres,G., Vaiana,G.S., Golub,L., and Rosner,R. 1978, ApJ 243, 288}
\refer{Serio,S., Reale,F., Jakimiec,J., Sylwester,B., and Sylwester,J. 1991, AA 241, 197.}
\refer{Warren,H.P., Winebarger,A.R., and Mariska,J.T. 2003, ApJ, (subm.)}
\refer{White,S.M., Thomas,R.J., Brosius,J.W., and Kundu,M.R. 2000, ApJ 534, L203}
\refer{Winebarger,A.R., Warren,H.P., and Seaton,D.B. 2003, ApJ, (subm.)}

\clearpage


\begin{deluxetable}{rrrrrrrrrr}
\footnotesize
\tablecaption{Cooling delays and iron abundances inferred from
	observations with TRACE ([A]=Aschwanden et al. (2000),
	[S]=Schrijver (2001), and [W]=Winebarger et al. (2003).}
\tablewidth{0pt}
\tablehead{
\colhead{No.}&
\colhead{Loop}& 
\colhead{Flux}& 
\colhead{Width}& 
\colhead{Length}&
\colhead{Time}&
\colhead{Electron}&
\colhead{Iron}&
\colhead{Radiative}&
\colhead{Conductive}\\
\colhead{}&
\colhead{}& 
\colhead{}& 
\colhead{}& 
\colhead{}&
\colhead{delay}&
\colhead{density}&
\colhead{abundance}&
\colhead{cooling}&
\colhead{cooling}\\
\colhead{}&
\colhead{\#}& 
\colhead{$F_{171}$ [DN/s]}&
\colhead{w [Mm]}&
\colhead{L [Mm]}&
\colhead{$\Delta t$ [min]}&
\colhead{$n_e$ [$10^9\ cm^{-3}$]}&
\colhead{${\alpha}_{Fe}$}&
\colhead{${\tau}_{rad}$ [min]}&
\colhead{${\tau}_{cond}$ [min]}}
\startdata
 1 & \#8 [A]	& 23.30	& 6.1	& 19.4	 & 2.53	& 3.2 & 4.2	& 6.7	& 141.2	\\
 2 & \#10 [A]	& 38.97	& 2.2	&  6.0 	 & 1.12	& 7.0 & 3.0	& 3.0	&  29.1 \\
 3 & \#12 [A]	& 26.84 & 2.3	&  5.8 	 & 3.01	& 5.6 & 2.2	& 8.0 	&  22.1	\\
 4 & \#16 [A]	& 21.08 & 3.5	&  9.9 	 & 1.76	& 4.0 & 4.0	& 4.7	&  46.2 \\
 5 & \#20 [A]	& 19.67 & 5.1	& 15.9	 & 0.82	& 3.2 & 7.4	& 2.2	&  95.3 \\
 6 & \#55 [A]	& 14.19 & 7.2	& 14.5	 & 0.61	& 2.3 & 12.0	& 1.6	&  56.7 \\
 7 & \#73 [A]	& 14.00 & 3.2	& 12.7 	 & 0.89	& 3.5 & 6.7	& 2.4	&  64.8 \\
 8 & \#190 [A]	& 15.41 & 2.9	&  7.4 	 & 0.76	& 3.8 & 6.6	& 2.0	&  24.2 \\
 9 & \#256 [A]	& 14.69 & 1.8	&  2.9   & 0.86	& 4.7 & 5.0	& 2.3	&   4.6 \\
10 & \#315 [A]	& 15.82 & 2.9	& 13.1   & 1.61	& 3.9 & 4.5	& 4.3	&  76.9 \\
11 & \#380 [A]  & 16.28 & 1.8	&  2.9   & 0.68	& 5.0 & 5.3	& 1.8	&   4.9 \\
12 & \#21 [S]	&  $-$  & $-$	& 32.0	 & 4.2	& $-$   & $-$	&   $-$ \\
13 & \#54 [S]	&  $-$  & $-$	& 33.0	 & 4.2	& $-$   & $-$   &   $-$ \\
14 & \#86 [S]	&  $-$  & $-$	& 41.0	 & 4.2	& $-$   & $-$   &   $-$ \\
15 & \#120 [S]	&  $-$  & $-$	& 45.0	 & 3.3  & $-$   & $-$   &   $-$ \\
16 & \#1 [W]	& 38.24 & 2.2	& 13.0	 & 2.50	& 6.9 & 2.0	& 6.6	&  135.2 \\
17 & \#2 [W]	&  8.85 & 3.2	& 65.0	 & 23.3	& 2.8 & 1.6	& 61.8	& 1350.0 \\
19 & \#3 [W]	& 36.69 & 5.5	& 102.0	 & 23.3	& 4.3 & 1.1	& 61.8	& 5160.0 \\
18 & \#4 [W]	& 30.48 & 3.0	& 78.0	 & 10.0	& 5.3 & 1.3	& 26.5  & 3720.0 \\
20 & \#5 [W]	&  9.16 & 12.0	& 178.0	 & 183.0 & 1.4 & 1.1  & 485.8 & 5310.0 \\
\enddata
\end{deluxetable}

\clearpage


\begin{figure} 
\plotone{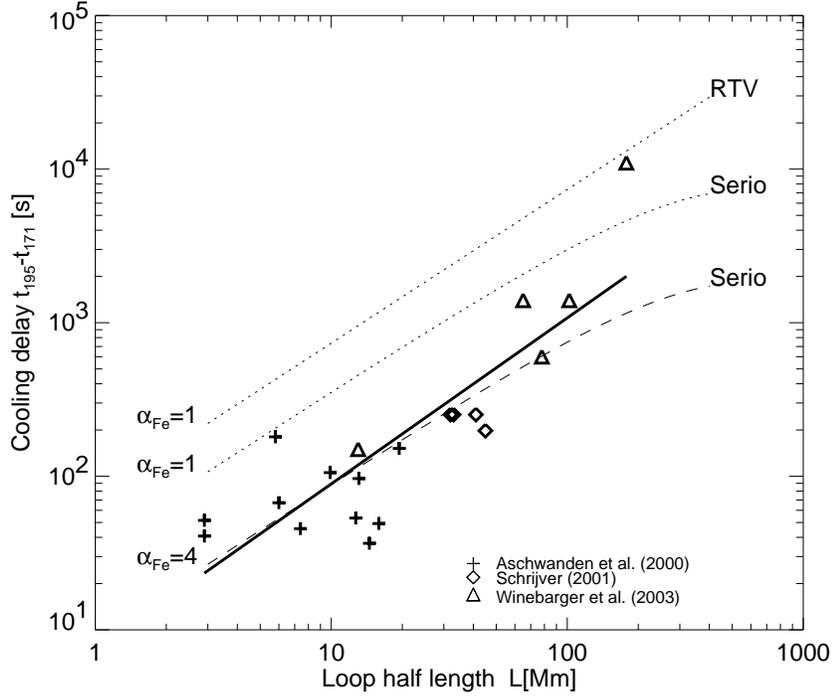}
\caption{Cooling delays $\Delta t_{12}$ are measured between the peak time in the
TRACE 195 \ang\ and 171 \ang\ filters, as function of the loop half length $L$,
from three datasets with 11 nanoflare loops (crosses; Aschwanden et al. 2000),
4 active region loops (diamonds; Schrijver 2002), and 
5 active region loops (triangles; Winebarger et al. 1003). The thick line represents
a linear regression fit with a slope of $1.08\pm0.16$. The theoretically predicted
scaling laws (based on RTV and Serio et al.) are shown for an iron enhancement
factor of ${\alpha}_{Fe}=1.0$ (dotted) and ${\alpha}_{Fe}=4.0$ (dashed).}
\end{figure}

\begin{figure} 
\plotone{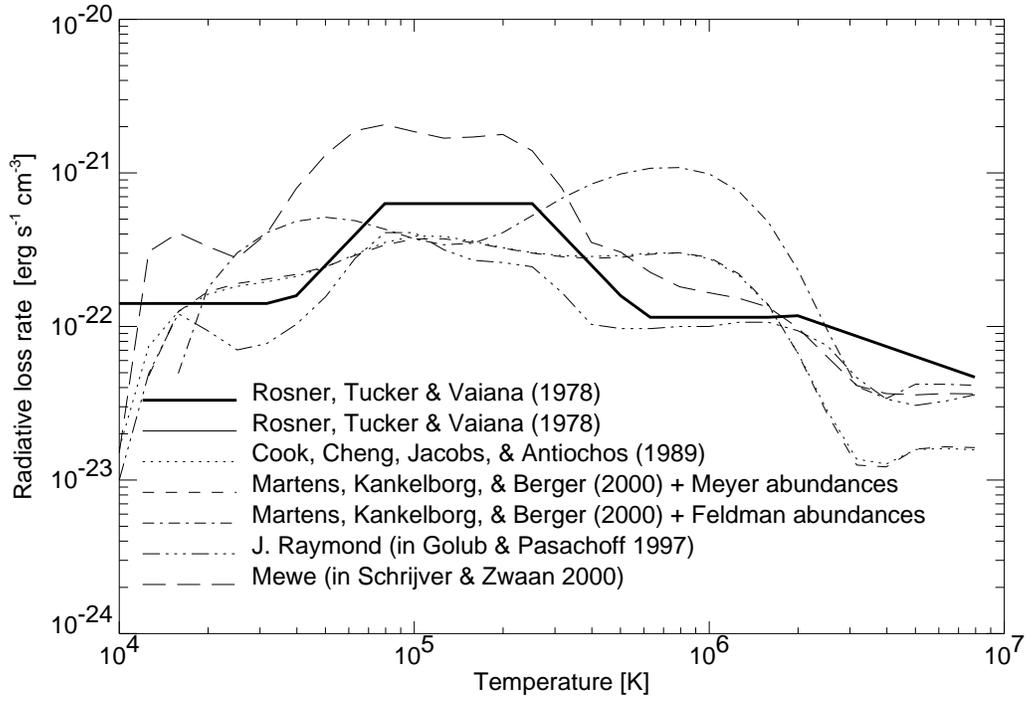}
\caption{
A compilation of radiative loss functions is shown. The differences mainly
result from the assumptions of elemental abundances. Coronal abundances (e.g. Feldman 1992)
have a $\approx 3$ times higher iron content than photospheric abundances (e.g. Meyer 1985),
and thus increases the value of the radiative loss function by the same factor at
temperatures around $T \approx 0.5-2.0$ MK. The one-piece powerlaw approximation is used
in the derivation of the RTV scaling law.}
\end{figure}

\end{document}